\documentclass{article}

\usepackage{arxiv}

\usepackage[utf8]{inputenc} 
\usepackage[T1]{fontenc}    
\usepackage{hyperref}       
\usepackage{url}            
\usepackage{booktabs}       
\usepackage{amsfonts}       
\usepackage{nicefrac}       
\usepackage{microtype}      
\usepackage{lipsum}		
\usepackage{graphicx}
\usepackage[numbers]{natbib}
\usepackage{doi}
\usepackage{float}
\usepackage{amsmath}

\renewcommand{\headeright}{}
\renewcommand{\undertitle}{}

\title{\textbf{E-Commerce in Africa: Divergent Impacts on Rural and Urban Economies} }

\date{}

\author{
  Jaelyn S. Liang \thanks{Corresponding Author } \thanks{Acknowledgments: Sriramteja Yerneni, Sowmiya D. Raja, Vedant Vohra, Alin Jain. } \\
  \texttt{liangjaelyn0999@gmail.com} \\
  \And
  Rehaan S. Mundy \\
  \texttt{rehaansmundy@gmail.com} \\
  \AND
  \centerline{Shriya Jagwayan} \\
  \centerline{\texttt{shriyajpr99@gmail.com}} \\
}

\hypersetup{
pdftitle={E-Commerce in Africa: Divergent Impacts on Rural and Urban Economies},
pdfsubject={q-bio.NC, q-bio.QM},
pdfauthor={Jaelyn L, Rehaan M, Shriya J},
pdfkeywords={First keyword, Second keyword, More},
}

\begin{document}
\maketitle

\begin{abstract}
	E-commerce is rapidly transforming economies across Africa, offering immense opportunities for economic growth, market expansion, and digital inclusion. This study investigates the effects of e-commerce on select African regions. By utilizing readiness factors, including mobile money deployment, GDP per capita, internet penetration, and digital infrastructure, the preparedness of African countries for e-commerce adoption is quantified, highlighting significant disparities. Through case studies in urban and rural areas, including Lagos, Kano, Nairobi, and the Rift Valley, the study shows e-commerce's significant effects on small and medium-sized enterprises (SMEs), employment, and market efficiency. Urban centers demonstrated significant gains in productivity and profitability, whereas rural regions experienced slower growth due to limited internet access and infrastructural barriers. Despite these challenges, localized solutions such as mobile money systems and agricultural e-commerce platforms are bridging gaps. This study highlights the significant potential of e-commerce in Africa while emphasizing the need for targeted investments and strategies to address existing regional disparities. 
\end{abstract}

\section{Introduction}
The urban-rural divide in African economies is an important factor influencing the growth and development of e-commerce across the continent. In Nigeria, urban economies like Lagos are thriving hubs of digital activity with advanced e-commerce infrastructure, internet access, and robust financial systems. Conversely, internet and digital access remain limited in rural areas such as Northern Nigeria. In the same way, in Kenya, urban places such as Nairobi are thriving in the e-commerce sector, having high mobile penetration and technological innovation. However, rural regions like Western Kenya face problems with infrastructure and lower internet connectivity, restricting their access to the e-commerce industry. Looking towards South Africa, cities like Johannesburg and Cape Town benefit from developed infrastructure and digital payment systems, but rural economies like Eastern Cape and Limpopo struggle with connectivity issues and online payment systems (Finance Corporation 5). This urban-rural divide in African economies highlights the need for targeted policies to enhance digital literacy and accessibility in rural areas to ensure that e-commerce growth impacts a larger portion of the population and contributes to more economic development across the continent (Telecommunication Union 40).

Due to rising poverty rates and economic disparities within Sub-Saharan Africa (S-SA), the UN’s Sustainable Development Goal 1 (SDG 1) to end poverty by 2030 will not be met (Cordes and Marinova). Fortunately, due to the rapid growth of Information Technology (IT) within the last three decades, policymakers can look to electronic commerce (EC) as a promising pathway for poverty alleviation (Arya et al.) (Devi). EC adoption contributes to overall poverty reduction by allowing for market diversification, greater access to consumers, and lowered costs of goods (Arya et al.). In one study alone, Kulshrestha and Saini found that 50\% of the population engaged with online shopping due to its easy accessibility. Due to the conveniences accompanying EC, Haji finds that the growth of EC within the countries of Brazil, Russia, India, China, and South Africa (BRICS) has led to the largest poverty reduction each nation has seen within the last 30 years. This was particularly true in their rural and remote regions, where EC aided in poverty alleviation by allowing for access to cheap products, job acquisition, and growth of MSMEs (Haji). Thus, considering Cordes and Marinova’s conclusion on the growing rate of poverty in Sub-Saharan Africa, EC is considered an important factor in developing the economic future of the region. 

E-commerce refers to the buying or selling of goods over the Internet. Through E-commerce, users can find a wide range of products and services. Out of the many forms of E-commerce present on our planet, the most prominent are digital marketplaces, online retail, and mobile commerce (“A Beginner’s Guide to E-Commerce”). As of now, e-commerce has seen significant growth globally, primarily driven by advances in technology, the addition of smartphones, and the increasing utilization of digital payment systems (Klenow). Rural economies are predominantly defined by their reliance on agriculture, small-scale industries, and limited infrastructure (Saiesha). Furthermore, these areas usually have low population densities and face challenges such as limited access to technology, and insufficient economic opportunities (Blasio). Urban economies are renowned for their higher population density, diverse industries, and strong access to technology. These factors contribute to more dynamic economic activities making urban economies the early adopters of e-commerce (Schöder et al.). Additionally, Urban Economies benefit from robust internet access and higher monetary income, ultimately facilitating the rapid growth of online business activities (Blasio). The adoption of E-commerce in these economies varies drastically. Urban areas, with their superior infrastructure and greater technological knowledge, have embraced E-commerce more quickly. In contrast,  rural economies, with their limited technological penetration and limited access to digital platforms, have been slower to adapt (Saiesha). 

Implementing e-commerce in rural areas is especially hard due to the lack of resources, gaps in knowledge, and low affordability. Rural areas lack basic resources such as efficient and competent internet infrastructure that are integral to developing rural e-commerce. Furthermore, rural areas tend to be lacking in digital literacy and other necessary skills that are essential to e-commerce. Moreover, due to the low affordability generally seen in rural regions, rural consumers have less purchasing power and rural producers cannot fulfill large orders with high-quality goods (Chatterjee 1, 4). However, if rural areas do manage to overcome these issues, rural producers and consumers will gain many advantages. Producers will have access to the global market, the ability to easily scale their businesses, and low-cost advertising. Consumers will gain 24/7 access to lower prices, a global marketplace, and a better shopping experience (Okolie and Ojomo 9-10). These benefits will help rural areas by improving the quality of life and catalyzing economic growth.

This research offers policymakers the knowledge and the means to help them understand how to make policies that help everyone, especially in places that are different, like cities and the countryside in Africa. It encourages fair economic growth for everyone. For businesses, the study shows that there are many opportunities in places that are not well-served, like rural areas, and that it would be beneficial if they make plans to work with these places for long-term success. This paper is aimed at exploring the different impacts of e-commerce on rural and urban economies in African countries. Our study begins with an introduction that establishes our sources, and key concepts and highlights significant aspects of the study. It then reviews existing literature specifically analyzing key factors influencing e-commerce growth in both rural and urban areas, discusses theories and models, and identifies gaps. We examine factors influencing e-commerce growth in both urban and rural areas. 

This paper provides an analysis of e-commerce’s differential impacts on rural and urban economies in African countries. We present a data-driven framework for assessing e-commerce readiness across diverse economies using standardized metrics and clustering techniques, identifying key factors that drive growth and gaps that hinder development. Our empirical analysis employs comparative case studies of Nigeria, Kenya, and South Africa to evaluate the role of infrastructure, digital literacy, and government policies in shaping e-commerce adoption. By integrating insights from statistical analysis, economic indicators, and existing theories, we explore strategies to bridge the urban-rural divide and promote equitable economic growth.

\section{Literature Review}
Digitization within the 21st century has paved the way for global e-commerce (EC) growth (Ahi et al). With the growing presence of EC internationally, it is critical to understand the benefits and costs accompanying its adoption. In terms of benefits, EC offers low-cost entry of goods into the international market and rapid response to demand fluctuations (Ahi et al). For businesses, small and medium sized enterprises especially, this easily allows them to tap into a more diverse market while retaining the power to easily adapt in times of demand changes. However, EC is not without shortcomings (Ahi et al). First, cross-border EC makes companies susceptible to data theft and privacy issues as well as various political, legal, and social risks. Second, the lack of necessary transparency has led to ambiguous returns, jeopardizing the profit the company makes. Moreover, it is also important to note that SMEs lag behind larger firms in actually adopting EC, leading to a major gap in EC adoption among developed and lesser developed nations. Regardless, the potential benefits brought by EC must be analyzed in the context of its adoption in poverty-stricken nations, such as Nigeria, Kenya, and South Africa.

E-commerce provides a cheaper, more modern way of selling and buying goods and has the potential for large growth in African urban economies. However, it has lagged due to being hindered historically by the high prices and poor quality of internet services, as well as the reliability and cost of delivery mechanisms (Goga et al.). The Economic Commission for Africa, through its African Information Society Initiative (AISI), has identified e-commerce as one of the four key areas in Africa to exploit ICTs to best advance social and economic development (Esselaar and Miller 19). The opportunities for the pathways commerce can provide are endless. A paper by Michael-Onuoha et al suggests that libraries can provide many opportunities for African countries, including ICT skills development, mobile information services, consulting, and agricultural information services to contribute to sustainable development (-Onuoha et al). However, a large contrast between infrastructure in African countries and developed regions has led to countries being unable to adopt e-commerce and showing a trend in its telephone density. “Telecommunications statistics from 1988 show a sharp contrast in the telephone line infrastructure and telephone density (teledensity) existing between the industrialized regions of Europe, North America, Asia, and Oceania and the developing regions of Africa, Asia, and Latin America (Ndonga). In an ICT (Information and co.. technology) study “Urban slums in Nairobi, Kenya, Wamuyu learned that few households had access to the Internet due to the inaccessibility of public Internet access points (Cordes and Marinova). A culmination of unaffordable services and services needing maintenance limit the usefulness of digital amenities. Furthermore, a lack of digital literacy similarly restricts low-income urban economies' use of e-commerce. Cybersecurity issues and online fraud have also been a prevailing reason why African countries haven't been effectively transitioning to means of e-commerce. While countries have attempted to place cybercrime security measures “, the presence of cybercrime legislation in these few countries has not guaranteed the efficient protection of online users within these jurisdictions (Ndonga). 

E-commerce growth in urban areas is stimulated by several key factors: high population, advanced infrastructure, and the busy lifestyle associated with urban life. The high population in urban areas offers a vast and diverse consumer base for online businesses. This increases both the sales of products and the variety of products needed to fulfill the needs of all residents. Additionally, the advanced infrastructure seen in urban areas, which contributes to both the reliable internet and intricate logistical support systems, along with the recent addition of AI facilitates a better experience for customers with tailored browsing experiences and rapid delivery (PLOS ONE, 2024). Moreover, due to its online nature, e-commerce fits perfectly into the prevalent usage of smartphones and digital payment platforms (Zhu, Li, \& Wang, 2023). Together, these reasons contribute to the drive towards e-commerce seen in urban areas. 

The adoption of e-commerce in developing economies in Africa can be understood through several theoretical frameworks. The Technology Acceptance Model (TAM) suggests that perceived usefulness and perceived ease of use are critical factors influencing e-commerce adoption (Davis, 1989). The TAM framework highlights how the perceived benefits of e-commerce, such as increased market reach and operational efficiency, can drive adoption, though challenges like internet infrastructure and digital literacy remain barriers (Molla \& Licker, 2005). The Diffusion of Innovations (DOI) theory, proposed by Rogers (2003), emphasizes the role of perceived attributes of innovations, such as relative advantage and compatibility, in the adoption process. Research indicates that in Africa, the relative advantage of e-commerce, such as cost reduction and expanded customer base, is significant, but issues related to compatibility with existing business practices and technological infrastructure can hinder adoption (Oliveira \& Martins, 2011). Additionally, the Institutional Theory underscores how formal and informal institutions, including government policies and cultural norms, affect e-commerce adoption. In Africa, supportive policies and initiatives by governments and NGOs are crucial in fostering an environment conducive to e-commerce growth, while institutional voids and regulatory challenges present obstacles (Zhu et al., 2006). These models collectively provide a framework for understanding the multifaceted nature of e-commerce adoption in developing African economies, revealing both the drivers and barriers of technology integration.

The increase in the use of e-commerce in Africa displays its potential to drive economic growth and development. However, by conducting an extensive literature review, we find gaps in the existing literature regarding the impact on urban vs. rural areas. While browsing the existing literature, we found that there is a lot more concentration on modern and urban areas compared to rural regions, disregarding the challenges and areas of growth that rural inhabitants face and experience. In addition, rural areas have big disadvantages compared to urban areas, namely, the lack of access to electronic and internet technologies, restrictive laws, and lower digital literacy rates. Existing literature briefly touches upon such downsides, but there is no mention of their effects on the e-commerce industry. Limited studies have been done on the differences in lifestyle in rural regions. This glosses over the subtle details of e-commerce’s effect, so more research needs to be done on opportunities that e-commerce gives to rural inhabitants. These are some of the many limitations that e-commerce poses and are some of the pressing issues we can address.

\section{Methodology}
\label{sec:headings}
The overall research approach for this study consists of two parts: (1) a data-driven analysis utilizing existing datasets and reports to calculate and compare e-commerce readiness across African countries, and (2) a comparative case study approach focused on Nigeria, Kenya, and South Africa. The first part involves leveraging standardized quantitative metrics to assess readiness across diverse economies, while the second part includes a thorough analysis of existing literature, government reports, industry analyses, and databases to explore the impact of e-commerce on both rural and urban economies in these countries. By examining diverse e-commerce environments—Nigeria's rapidly expanding market, Kenya's mobile money-driven adoption, and South Africa's advanced infrastructure, alongside a broader readiness analysis across African countries, the study aims to comprehensively compare different paths to e-commerce adoption. 

\subsection{E-commerce Readiness Analysis}
Analysis on e-commerce readiness evaluates the capacity of African countries to adopt and scale e-commerce systems based on quantitative metrics. Here, we employ an approach to measure and compare readiness levels across multiple African countries.  
\subsubsection{Data Collection}
Calculation of readiness scores for each country utilized features selected from publicly available datasets and reports, such as GSMA, Statistica, World Bank Group, and Worldometer. These factors include

\begin{enumerate}
    \item GDP per Capita: Higher GDP per capita typically indicates a population with greater disposable income, something essential for e-commerce adoption. \cite{worldbank_gdp}
    \item Mobile Money Deployments: Measures the number of mobile payment services available in the country, indicating financial infrastructure readiness, especially in areas where traditional banking methods are scarce. \cite{gsma_mobile_money}
    \item Internet Penetration (\%): Indicates the percentage of population with access to the internet, which is a fundamental requirement for e-commerce. \cite{statista_internet_penetration}
    \item Population Size: A proxy for market potential, larger populations represent larger potential customer bases for e-commerce platforms. \cite{worldometers_population}
\end{enumerate}

From here, a dataset was created, holding countries as rows and the selected factors as columns. This dataset serves as a foundation for subsequent analysis, allowing for the calculation of e-commerce readiness scores across countries.  
\subsubsection{Data Normalization}
As the selected features are measured on different scales, normalization was performed to standardize the data. Each feature was scaled to a range between 0 and 1 using Min-Max Normalization. With \(\text{Value}\) as the original data point for a feature and \(\text{Min Value}\) and \(\text{Max Value}\) as the minimum and maximum values of the feature across all countries, we compute this as:

\[
\text{Normalized Value} = \frac{\text{Value} - \text{Min Value}}{\text{Max Value} - \text{Min Value}}
\]

This ensures no feature dominates the calculation due to its magnitude. 

 \subsubsection{Weight Assignment}
 Weights were assigned to each feature based on its relative importance to e-commerce readiness. The weights are as follows:

\begin{enumerate}
    \item GDP per Capita: 30\% (0.3)
    \item Mobile Money Deployments: 40\% (0.4)
    \item Internet Penetration (\%): 20\% (0.2)
    \item Population Size: 10\% (0.1)
\end{enumerate}
These values are a reflection of the importance of financial infrastructure (mobile money), economic development (GDP per capita), and connectivity (internet access) in enabling e-commerce, while accounting for the scale of market potential (population).

\subsubsection{Readiness Scores}
The Readiness Score was computed as a weighted sum of the normalized feature values:

\begin{align*}
\text{Readiness Score} &= (W_1 \times \text{GDP per Capita (Normalized)}) \\
&\quad + (W_2 \times \text{Mobile Money Deployments (Normalized)}) \\
&\quad + (W_3 \times \text{Internet Penetration (Normalized)}) \\
&\quad + (W_4 \times \text{Population (Normalized)})
\end{align*}

Where \(W_1\), \(W_2\), \(W_3\), and \(W_4\) are the weights assigned to each feature.

Scores range from \textbf{0 to 1}. A score close to \textbf{1} indicates high readiness for e-commerce, driven by robust infrastructure, digital engagement, and market potential. A score close to \textbf{0} suggests significant challenges in terms of financial, digital, or market readiness. Data visualization methods were employed to display the readiness scores and highlight patterns across countries. Visualizations, such as heatmaps and bar charts, were generated in Google Colab using Python libraries, including Matplotlib and Seaborn. These visualizations were integral to identifying trends and comparing e-commerce readiness across regions. 
 
\subsection{Case Studies}
Data was obtained from Nigeria, Kenya, and South Africa for the case study analysis. These countries were chosen based on the availability of secondary data, the size of economies, and regulatory factors within the countries. Nigeria was found to be Africa’s largest economy, with a dynamic e-commerce market and significant mobile penetration. However, the country did face significant trust issues with payments and logistics. Meanwhile, Kenya has a strong mobile money infrastructure, growing urbanization, and rapid adoption of digital solutions. Lastly, South Africa has advanced infrastructure, high digital literacy, and well-established legal frameworks for data protection. In this study, we aim to explore how these various factors affect e-commerce adoption in rural versus urban areas within African economies.

To analyze the differing effects of e-commerce on rural and urban economies in Africa, we utilize secondary data chosen for relevance and credibility. Peer-reviewed academic research provides insights into the economic and social impacts of e-commerce (Aker \& Mbiti, 2010), while government publications contribute data on infrastructure, economic patterns, and digital literacy, contextualizing regional disparities (African Development Bank, 2020). Industry analyses further inform market trends and consumer behavior across the continent (Manyika et al., 2013). Additionally, databases such as those from the International Telecommunication Union (2021) and the World Bank are used to access detailed economic and demographic data critical to evaluating e-commerce growth.

This study compares urban centers like Johannesburg and Nairobi to rural regions such as Limpopo and Machakos to explore key issues, including internet accessibility, costs, and digital literacy disparities. The analysis investigates how variations in urbanization and infrastructure shape e-commerce adoption in different contexts. Using a combination of content, meta, statistical, and document analyses, we examine themes such as infrastructure development, internet penetration, regulatory frameworks, and economic indicators. Content analysis, supported by NVivo, helps identify patterns in government policies and industry reports, while meta-analysis synthesizes findings from multiple studies to highlight regional trends. Statistical analysis quantifies relationships between e-commerce growth and factors like GDP, internet penetration, and mobile usage, using tools such as R (Khan \& Uwemi, 2022). Document analysis further examines country-specific policies, offering insights into how national contexts influence adoption strategies (Olusola, 2022).

\section{Results}
\subsection{E-Commerce Readiness Analysis}
The analysis of e-commerce readiness scores highlights disparities across African countries, reflecting their varying levels of infrastructure, financial inclusion, and digital access. Through data visualization and analysis, we identify patterns that highlight the leading and lagging countries in e-commerce readiness and display regional trends across the continent.  
\subsubsection{Readiness Scores}

\vspace{-0.45cm} 
\begin{figure}[H]
    \centering
    \includegraphics[width=0.6\textwidth]{"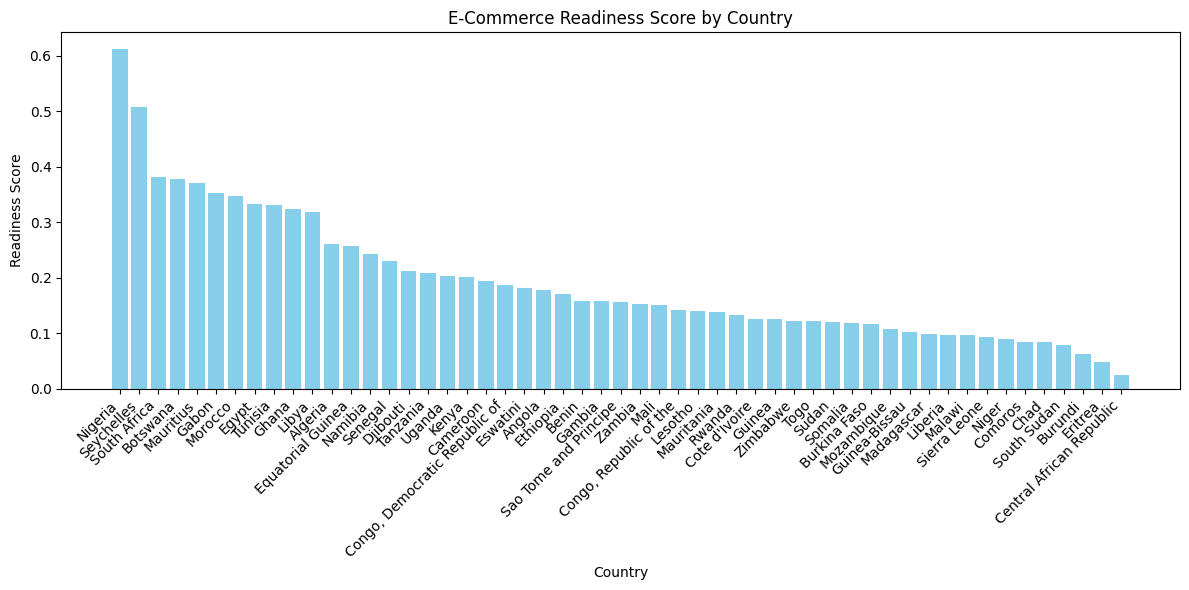"}
    \caption{\textbf{E-Commerce Readiness Score by Country. }The bar graph shows the e-commerce readiness score for each of the African countries, with higher scores indicating greater readiness.}
    \label{fig:ecommerce_readiness}
\end{figure}
\vspace{-0.4cm} 

\par

The readiness scores, ranging from 0.025  to 0.611, show a wide spectrum of preparedness for e-commerce adoption. Countries with high scores, such as Nigeria, South Africa, and Kenya, have strong infrastructure, digital connectivity, and large mobile money availability. Conversely, nations like Chad and the Central African Republic rank lower due to challenges in internet penetration, financial services, and market access. These results emphasize the role of enabling factors in e-commerce growth.  
\begin{figure}[H]
    \centering
    \includegraphics[width=0.6\textwidth]{"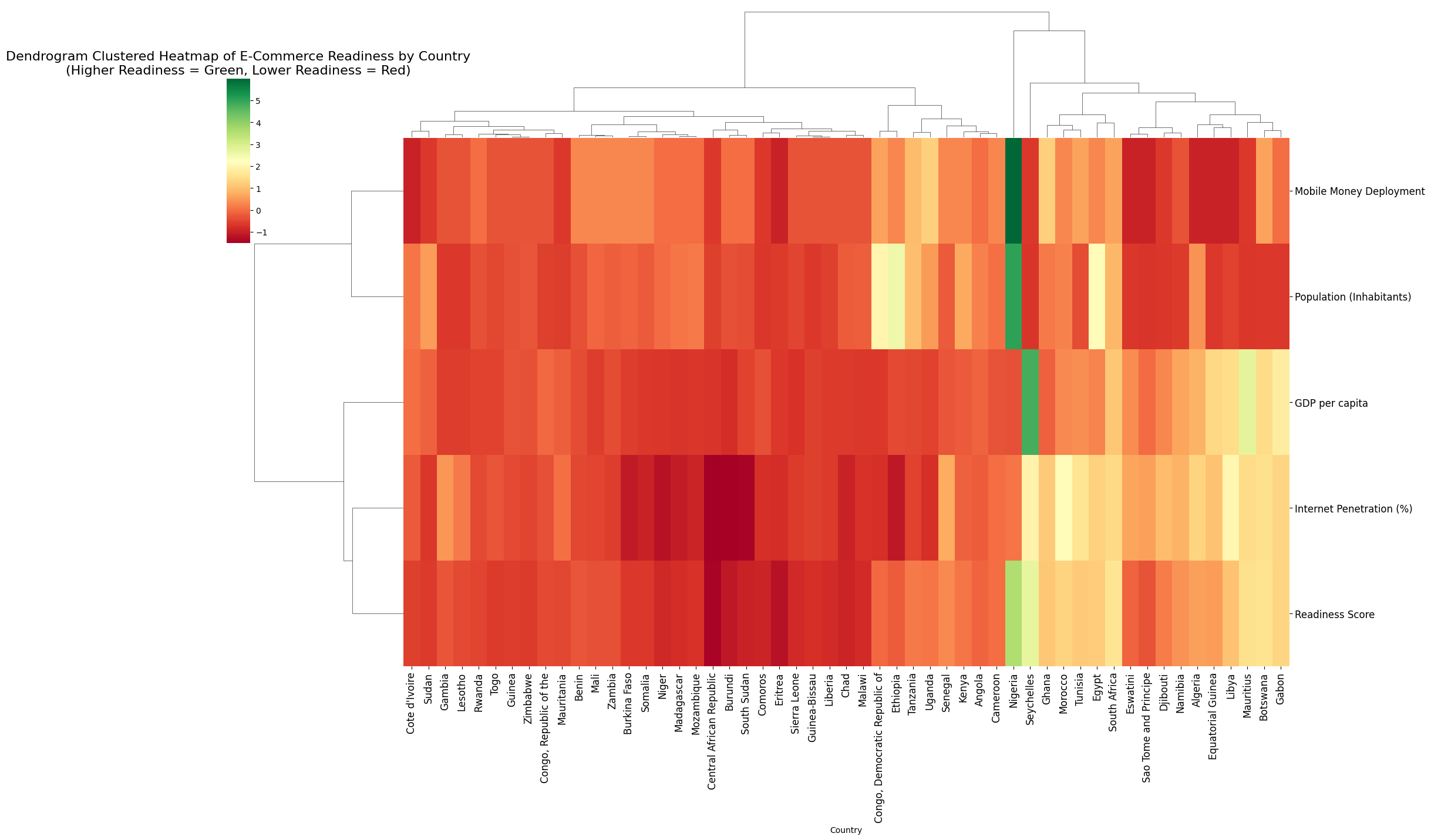"}
    \caption{\textbf{Dendrogram \textbf{Clustered Heat map of E-Commerce Readiness Scores across African Countries.} }The heat map visualizes the normalized values for GDP per capita, mobile money deployments, internet penetration, population size, and the computed readiness scores. The dendrogram on the left groups features based on their correlation. Features that are closely clustered (ex. GPD per capita and readiness score) have a strong relationship. The dendrogram on the top clusters countries with similar e-commerce readiness profiles. Higher readiness is represented by green shades, while lower readiness is shown in red.  }
    \label{fig:dendrogram_ecommerce_readiness}
\end{figure}
\vspace{-0.4cm} 
Fig 2 illustrates the exact relationship between African countries based on their readiness scores and enabling factors, like GDP per capita, internet penetration, population size, and mobile money deployments. Countries clustered together have similar readiness profiles and comparable levels of infrastructure, financial inclusion, and digital access. High-readiness countries like South Africa, Nigeria, and Kenya are in close proximity towards the right, while low-readiness countries, including the Central African Republic, Chad, and Burundi cluster towards the left. Medium-readiness countries, such as Ghana, Tunisia, and Morocco, form a mix of strengths and areas for improvement. By identifying such gaps, strategies for fostering balanced e-commerce growth can be created. 

\subsection{Case Studies}
E-commerce is constructing its place even in the northern states of Nigeria, particularly in the regions of Kano and Kaduna; however, relative development has yet to be achieved. Platforms like Jumia and Konga help farmers reach urban consumers directly, reducing reliance on brokers and increasing profitability by as much as 30\% (Adegboye, 2021). Unfortunately, rural areas are at a disadvantage due to factors such as constrained mobile internet access—only about 26\% of rural Northern Nigerians have reliable internet coverage (Nigerian Communications Commission, 2020)—and poor levels of computer literacy, with digital literacy rates hovering at just 20\% in some rural areas, affecting the uptake of digital platforms. In some rural areas, mobile money systems such as Paga have enabled over 10 million users to participate in e-commerce even without traditional banking facilities (Aker \& Mbiti, 2010). Nevertheless, structural problems such as an erratic power supply—where only 40\% of rural areas have consistent access to electricity—and poor transportation links remain challenges to the growth of e-commerce in these regions. Increasing internet access and improving digital literacy will be essential for unlocking the full potential of e-commerce in Kano and Kaduna.

In Lagos, Nigeria, e-commerce has transformed small-scale businesses by significantly enhancing their operations and market reach. This impact is primarily observed through various e-commerce structures like Business-to-Business (B2B) and Business-to-Consumer (B2C) models, which have streamlined transactions and interactions between businesses and consumers. Specifically, 62.3\% of businesses acknowledged that B2B structures have improved transactions, fostering better relationships between businesses, while 65.1\% highlighted that B2C structures have facilitated easier access to products for end-users. The adoption of e-commerce has led to quicker and more efficient transactions, reducing delays commonly associated with traditional commerce methods. It has also enhanced communication with partners and clients, allowing businesses to operate more fluidly. Additionally, e-commerce has enabled small businesses to reduce operational costs, increase branding efficiency, and use more effective marketing strategies. Despite challenges like high implementation costs, limited technological awareness, and security concerns, the overall effect of e-commerce on small-scale businesses in Lagos is positive. It has enabled these businesses to scale up operations, increase their market presence, and improve profitability. The case study (Labanauskaite) illustrates that e-commerce serves as a vital tool for economic growth and development, especially in urban areas like Lagos, where market dynamics can support digital transformation. 

E-commerce has significantly impacted the urban economy of Nairobi, Kenya, especially in the Small and Medium Enterprises (SME) sector. A study conducted by Kabuba reported that 66\% of SMEs declared an increase in sales due to the adoption of e-commerce (Brookings). Moreover, 60\% of SMEs experienced a decrease in overall costs of production due to an increase in logistical efficiency (Brookings). Furthermore, with the introduction of e-commerce into Nairobi there was an increase of 40\% in the employment rates, with digital companies creating new job opportunities (Brookings). However, there remain some infrastructural challenges with 35\% of digital businesses facing poor internet and 45\% dealing with the operational inefficiencies that come with poor transportation services (Brookings). Despite the aforementioned challenges, Nairobi continues to prove itself as a digital city, with e-commerce contributing to 10\% of its GDP and projected to account for more in the future (Brookings). This case study on Nairobi details the substantial benefits that have been created due to the adoption of e-commerce, however, it also highlights the need for infrastructural advancements, a recurring challenge in the implementation of e-commerce. 

E-commerce has changed rural economies in Kenya, most notably in the Rift Valley and Western Kenya where agriculture is many’s primary source of income. Platforms like M-Farm and Twiga Foods have transformed how farmers access markets by connecting them directly to urban buyers, allowing farmers to bypass the ‘middleman’ system and improve their profit margins (Aliu, 2024).  In addition, with the employment of these e-commerce platforms farmers are granted access to real-time information and market prices, allowing them to make more strategic decisions when transacting their goods (Morepje., et al. 24). Moreover, with the implementation of payment systems like M-Farm have been able to further streamline transactions, fostering financial transparency within the farmers (Baumüller, 2017). Furthermore, these digital platforms not only enhance market access but also contribute to fairer pricing for produce and other agricultural products, allowing the farmers to generate more revenue as well as reduce post-harvest losses  (Osten, 2021). 

South Africa, particularly in urban centers like Johannesburg and Cape Town is experiencing significant transformations in their economies due to the rise of e-commerce. The increasing internet and mobile usage have led to a surge in online shopping, creating new opportunities for local businesses and startups (Mamima, 2020). In Johannesburg, e-commerce has expanded the retail and service sectors, helping small businesses reach larger markets without physical “brick-and-mortar” storefronts (Worku, 2024). Similarly, Cape Town’s strong infrastructure has made it a hub for e-commerce startups, further driving innovation and employment in areas like warehousing and delivery services (Goga et al., 2019). However, challenges such as digital exclusion and income inequality persist, limiting the full impact of e-commerce in these urban economies (Wang, 2023).  

A thorough analysis of South African rural economies suggests that, with the adoption and implementation of proper digital infrastructure, e-commerce can positively affect the growth of SMEs (Siwundla) (Madzvamuse et al.) (Chiliya et al.). In South Africa, that is crucial because SMEs are the most important contributors to the economy, accounting for 30\% of the GDP and 80\% of the employment rate (Siwundla). Specifically, e-commerce can allow SMEs to enhance their productivity by allowing business transactions to happen seamlessly from different regions, increase their revenue by market diversification, and provide better customer support to consumers (Siwundla)(Chiliya et al.). Consequently, Siwundla finds that economies with e-commerce have seen increased consumer spending, reduced poverty, and overall economic growth. Unfortunately, in the status quo, there appears to be a lack of literature specifically analyzing the impact of e-commerce in South African rural economies due to the lack of SMEs' technological preparedness there (Madzvamuse et al.). Thus, although e-commerce holds great potential to revolutionize the SME sector, the lack of proper infrastructure prevents rural economies in South Africa, such as Limpopo and Eastern Cape, from fully benefiting (Madzvamuse et al.). 

\subsubsection{Empirical Analysis}
The statistical outcomes indicate that there is a noticeable variation in the influence that e-commerce has on rural economies as compared to urban economies in Africa. In metropolitan regions such as Lagos, 62.3\% of the respondents indicated that the transactions improved when conducting activities using Business-to-Business (B2B) models, while in Nairobi, e-commerce caused 66 percent of SMEs to make higher sales (Labanauskaite; Brookings). Such observations indicate the existence of an attractive urban center for the e-commerce business that enhances growing efficiency. On the contrary, in the northern region of Nigeria, Kano and Kaduna in particular, respondents reported that only 26\% of people had good and reliable internet access and that about 20\% of residents were computer literate (Nigerian Communications Commission, 2020). This reduces the impact of Jumia and Konga, which would increase farmers’ incomes by nearly thirty percent (Adegboye, 2021). Likewise, although M-Farm has been of great help to rural farmers in Kenya, the farmers still face several infrastructural constraints (Aliu, 2024). Global statistics corroborate these conclusions: several infrastructural and educational gaps contribute to rural areas growing at a lower rate as compared to urban areas. 
 
\section{Conclusion}

This study aimed to explore the effects of e-commerce in African regions and to develop a readiness score system that quantifies the potential for e-commerce growth. Through a thorough case study and an analysis of GDP per capita, mobile money deployments, internet penetration, and population, this research provides actionable insights into the factors enabling or hindering e-commerce adoption. 

\par The e-commerce readiness analysis, supported by a scoring model, showed significant disparities among African countries. High-scoring nations like Nigeria, Kenya, and South Africa demonstrated the impact of robust digital infrastructure, mobile money options, and economic growth on providing e-commerce ecosystems. Conversely, lower-scoring countries such as Chad and the Central African Republic underscored persistent challenges posed by inadequate infrastructure, limited financial inclusion, and low internet access. These emphasize the importance of addressing foundational gaps to foster economic growth across regions. 

\par Our case studies further contextualize these insights, showing how e-commerce has reshaped economies at both urban and rural areas. In cities like Lagos, Nairobi, Johannesburg, and Cape Town, e-commerce has boosted small and medium-sized enterprises (SMEs), increased employment opportunities, and enhanced operational efficiencies. These urban areas benefit from relatively advanced digital infrastructure and higher levels of digital literacy, facilitating smoother adoption of e-commerce platforms. Meanwhile, rural regions such as Kano, Kaduna, and the Rift Valley revealed the challenges of limited internet access, low digital literacy, and inadequate infrastructure, which hinder e-commerce's full potential despite localized successes like mobile money systems and agricultural platforms. The empirical analysis underscored the disparities in e-commerce impacts between urban and rural economies. While urban areas experienced many improvements in efficiency, profitability, and market access, rural areas continued to grapple with barriers that limit participation in the digital economy. 

\par Despite limitations, like data availability and scope, this research demonstrates the utility of readiness scores as a decision-making tool for stakeholders. By finding gaps and strengths, the model offers an option for targeted investments and policy reforms to bring e-commerce growth. Future work could include incorporating real-time data and expanding the scope to include cultural factors. 

\par Ultimately, this study highlights the transformative potential of e-commerce in Africa. By investing in digital infrastructure, and addressing rural-urban disparities, Africa can harness e-commerce as a tool to drive sustainable growth, reduce economic disparities, and create opportunities across the continent.

\end{document}